\documentclass{ieeetj}
\usepackage{cite}
\usepackage{amsmath,amssymb,amsfonts}
\usepackage{algorithmic}
\usepackage{balance}
\usepackage{url}
\usepackage{graphicx,color}
\usepackage{textcomp}
\usepackage{xcolor}
\usepackage{booktabs}
\usepackage{multirow,multicol}
\usepackage{hyperref}
\hypersetup{hidelinks=true}
\usepackage{algorithm,algorithmic}
\def\BibTeX{{\rm B\kern-.05em{\sc i\kern-.025em b}\kern-.08em
    T\kern-.1667em\lower.7ex\hbox{E}\kern-.125emX}}
\AtBeginDocument{\definecolor{tmlcncolor}{cmyk}{0.93,0.59,0.15,0.02}\definecolor{NavyBlue}{RGB}{0,86,125}}
\newcommand{\newpara}[1]{\vspace{4pt}\noindent\textbf{#1}}
\hypersetup{
    colorlinks=true,
    linkcolor=purple,
    filecolor=magenta,      
    urlcolor=purple,
}

\def\OJlogo{\vspace{-4pt}$<$Society logo(s) and publication title will appear here.$>$}
\def\seclogo{\vspace{10pt}$<$Society logo(s) and publication title will appear here.$>$}

\def\authorrefmark#1{\ensuremath{^{\textbf{#1}}}}

\begin{document}
\receiveddate{XX Month, XXXX}
\reviseddate{XX Month, XXXX}
\accepteddate{XX Month, XXXX}
\publisheddate{XX Month, XXXX}
\currentdate{XX Month, XXXX}
\doiinfo{XXXX.2022.1234567}

\markboth{}{Jung {et al.}}

\title{SpoofCeleb: Speech Deepfake Detection and SASV In The Wild}
\author{
Jee-weon Jung\authorrefmark{1}, Member, IEEE, 
Yihan Wu\authorrefmark{2}, Member, IEEE,
Xin Wang\authorrefmark{3}, Member, IEEE,\\
Ji-Hoon Kim\authorrefmark{4}, Member, IEEE,
Soumi Maiti\authorrefmark{5}, Member, IEEE,
Yuta Matsunaga\authorrefmark{6},\\
Hye-jin Shim\authorrefmark{1}, Member, IEEE, 
Jinchuan Tian\authorrefmark{1},
Nicholas Evans\authorrefmark{7}, Member, IEEE,\\
Joon Son Chung\authorrefmark{4}, Member, IEEE,
Wangyou Zhang\authorrefmark{8},
Seyun Um\authorrefmark{9}, Member, IEEE,\\
Shinnosuke Takamichi\authorrefmark{10}, Member, IEEE, 
Shinji Watanabe\authorrefmark{1}, Fellow, IEEE, 
}
\affil{Carnegie Mellon University, PA 15213 USA}
\affil{Renmin University of China, Beijing 100872 China}
\affil{National Institute of Informatics, Tokyo Japan}
\affil{Korea Advanced Institute of Science and Technology, Daejeon 34141 South Korea}
\affil{Meta, CA 94025 USA}
\affil{University of Tokyo, Tokyo 1130033 Japan}
\affil{EURECOM, Biot 06410 France}
\affil{Shanghai Jiao Tong University, Shanghai 200240 China}
\affil{Yonsei University, Seoul 03722 South Korea}
\affil{Keio University, Kanagawa 2238521 Japan}

\corresp{Corresponding author: Jee-weon Jung (email: jeeweonj@ieee.org).}
\authornote{Jee-weon Jung, Hye-jin Shim, Jinchuan Tian, and Shinji Watanabe utilized the Bridges2 system at PSC and the Delta system at NCSA through an allocation (CIS210014) from the Advanced Cyberinfrastructure Coordination Ecosystem: Services \& Support (ACCESS) program. This program is funded by National Science Foundation grants \#2138259, \#2138286, \#2138307, \#2137603, and \#2138296. Xin Wang is supported by JST, PRESTO Grant Number JPMJPR23P9, Japan. This work was partly supported by ONR N00014-231-2086.}

\begin{abstract}
This paper introduces SpoofCeleb, a dataset designed for Speech Deepfake Detection (SDD) and Spoofing-robust Automatic Speaker Verification (SASV), utilizing source data from real-world conditions and spoofing attacks generated by Text-To-Speech (TTS) systems also trained on the same real-world data. 
Robust recognition systems require speech data recorded in varied acoustic environments with different levels of noise to be trained. However, current datasets typically include clean, high-quality recordings (bona fide data) due to the requirements for TTS training; studio-quality or well-recorded read speech is typically necessary to train TTS models.
Current SDD datasets also have limited usefulness for training SASV models due to insufficient speaker diversity.
SpoofCeleb leverages a fully automated pipeline we developed that processes the VoxCeleb1 dataset, transforming it into a suitable form for TTS training. We subsequently train 23 contemporary TTS systems. SpoofCeleb comprises over 2.5 million utterances from 1,251 unique speakers, collected under natural, real-world conditions.
The dataset includes carefully partitioned training, validation, and evaluation sets with well-controlled experimental protocols. We present the baseline results for both SDD and SASV tasks. All data, protocols, and baselines are publicly available at \url{https://jungjee.github.io/spoofceleb}.

\end{abstract}


\maketitle

\section{INTRODUCTION}
\label{sec:intro}

\IEEEPARstart{T}{he} quality of synthetic speech has improved rapidly, driven by advancements in technologies such as flow matching, neural codecs, and speech-language modeling~\cite{lipmanflow,valle,chen2023vector}. These innovations have significantly enhanced the naturalness and intelligibility of generated speech. The increasing availability of open sources and APIs for Text-To-Speech (TTS) systems has made high-quality synthetic speech more accessible to the general public~\cite{ttsapi,ttsapi2}.

Although originally developed for positive applications, this technology is increasingly being exploited for malicious purposes~\cite{damiani2019voice,PM2024robocall}. Synthetic speech generated with harmful intent, often referred to as spoofing, is being used to deceive individuals in scenarios such as voice phishing (or vishing). 
Spoofing also undermines the reliability of speech biometric systems, including Automatic Speaker Verification (ASV), many of which remain highly vulnerable to such attacks~\cite{jung2024extent,wang2020asvs2019}.

In response to these challenges, several datasets have been developed to advance research in Speech Deepfake Detection (SDD)~\cite{wu2015sas,wu2015asvs2015,yi2022add2022}. 
For robust recognition systems, it is essential to have training data that cover a wide range of real-world acoustic environments and speaker diversity. 
However, speech generation systems, such as TTS and Voice Conversion (VC), typically require studio-quality or clean, read speech for training. 
Therefore, current datasets tend to feature clean, monotonic bona fide speech, with spoofed samples also being clean, as they are synthesized using TTS and VC systems trained on such data.
The emerging task of Spoofing-robust Automatic Speaker Verification (SASV)~\cite{jung2022sasv} lacks dedicated datasets.
Many SDD datasets also suffer from limited speaker diversity, which hinders research on SASV systems that require training with data from hundreds or even thousands of speakers.

To this end, we introduce SpoofCeleb, a dataset built upon VoxCeleb1~\cite{nagrani2017voxceleb}, a widely used ASV dataset consisting of the voices of $1,251$ celebrities recorded under real-world conditions. 
We also develop a fully automated pipeline that processes VoxCeleb1 to produce in-the-wild bona fide speech samples that can be used for training TTS systems.\footnote{The development of this pipeline is extensive, and the resulting bona fide speech data can serve other purposes, such as advancing research on TTS systems trained on noisy, in-the-wild data. We detail this aspect in a separate work, referring to the dataset as TTS In The Wild (TITW)~\cite{jung2024text}.}
From the two available TTS training sets in TITW, we use TITW-Easy as the source dataset to generate 23 spoofing attacks. SpoofCeleb is the first dataset explicitly designed for both SDD and SASV, where the bona fide speech is real-world, noisy speech. The dataset is divided into three subsets for training, validation, and evaluation, accompanied by evaluation protocols. 
Baseline systems trained on SpoofCeleb’s training set are also presented, demonstrating SpoofCeleb’s effectiveness in and potential for future research in SDD and SASV.

\begin{table}[t]
    \caption{
    List of datasets in Speech Deepfake Detection (SDD) and Spoofing-robust Automatic Speaker Verification (SASV).
    FakeAVCeleb~\cite{khalid2021fakeavceleb} and ``In The Wild~\cite{muller2022does}'' also have in-the-wild data. However, they have either only an evaluation set or the number of speakers or spoofing attacks is limited.
    }
    \label{tab:corpora_comparison}
    \centering
    \resizebox{\columnwidth}{!}{
    \begin{tabular}{l|cccc}
        \toprule
        Dataset & \# Spk & \# Utt & \# Attacks & Domain \\
        \toprule
        SAS~\cite{wu2015sas} & $106$ & $652,615$ & $9$ & studio-recorded \\
        ASVspoof2015~\cite{wu2015asvs2015} & $106$ & $263,151$ & $10$ & studio-recorded \\
        Noisy Datsbase~\cite{tian2016noisy} & $106$ & $263,151$ & $10$ & studio-recorded \\
        ASVspoof2019 LA~\cite{wang2020asvs2019} & $107$ & $121,461$  & $19$ & studio-recorded\\
        ASVspoof2021 LA~\cite{yamagishi2021asvs2021} & $67$ & $164,612$ & 13 & studio-recorded \\
        ASVspoof2021 DF~\cite{yamagishi2021asvs2021} & $93$ & $593,253$ & 100+ & studio-recorded \\
        Voc.v~\cite{wang2023vocv} & $21$ & $82,048$ & $8$ & studio-recorded \\
        PartialSpoof~\cite{zhang2022partialspoof} & $107$ & $121,461$ & $19$ & studio-recorded \\
        WaveFake~\cite{frank2021wavefake} & $2$ & $136,085$ & $9$ & studio-recorded \\
        ADD 2022~\cite{yi2022add2022} & N/R & $493,123$ & N/R & studio-recorded \\
        ADD 2023~\cite{yi2023add2023} & N/R & $517,068$ & N/R & studio-recorded \\
        HAD~\cite{yi2021had} & $218$ & $160,836$ & $2$ & studio-recorded \\
        CFAD~\cite{ma2023cfad} & $1302$ & $347,400$ & $12$ & studio-recorded \\
        ASVspoof5~\cite{wang2024asvs5} & $1,922$ & $1,004,081$ & 32 & audiobook \\
        MLAAD~\cite{muller2024mlaad} & N/R & $76,000$ & $54$ & mixed\\
        FMFCC-A~\cite{zhang2021fmfcca} & $131$ & $50,000$ & $13$ & N/R \\
        \midrule
        FoR~\cite{reimao2019for} & N/R & $195,541$ & $7$ & in the wild \\
        FakeAVCeleb~\cite{khalid2021fakeavceleb} & 500 & $11,857$ & 1 & in the wild\\
        In-The-Wild~\cite{muller2022does} & $58$ & $31,779$ & N/R & in the wild\\
        VSASV~\cite{hoangvsasv} & $1,382$ & $338,000$ & $3$ & mixed\\

        \textbf{SpoofCeleb} & \textbf{1,251} & \textbf{2,687,292} & \textbf{23} & \textbf{in the wild} \\
        \bottomrule
    \end{tabular}
    }
    \vspace{-0.7cm}
\end{table}





    %

\section{RELATED WORKS}
\label{sec:related}
\newpara{Datasets for SDD and the generation-recognition trade-off.}
To safeguard the authenticity of speech, several datasets have been published to support research in SDD~\cite{wu2015sas,wu2015asvs2015,tian2016noisy,wang2020asvs2019,wang2024asvs5,zhang2022partialspoof,yi2022add2022,yi2023add2023}. One of the most critical decisions when creating these datasets is the selection of the source data (i.e., bona fide speech). This decision involves a trade-off, which we refer to as the ``{\em generation-recognition trade-off}.''

For both SDD and SASV on the recognition side, incorporating data with diverse noise, reverberation, and varied domains is essential for training robust models. It is well known that recognition models trained solely on clean speech often struggle to effectively generalize to noisy environments during inference~\cite{tian2016noisy}. While data augmentation techniques can help mitigate this issue~\cite{Tak2021}, the most effective solution is to use training data drawn from a wide range of real-world sources.

Conversely, traditional TTS training requires a carefully curated and recorded dataset. Sentence prompts must be selected to ensure comprehensive phonetic coverage~\cite{kominek04b_ssw}, and recordings are typically made by voice professionals in clean environments, ideally in a single anechoic studio. These recordings are of high studio quality and carefully articulated but are not scalable. For instance, the well-known CMU Arctic database includes recordings from fewer than 10 voice professionals, each reading approximately 1,000 speech prompts~\cite{kominek04b_ssw}.
Modern TTS systems, however, often require significantly more training data. Instead of relying on these small-scale, TTS-specific databases, contemporary models frequently use audiobook datasets (e.g., MLS~\cite{pratap2020mls}), which, while not studio-grade, consist of relatively clean audiobook recordings made by numerous readers in their homes or offices.

Current SDD datasets tend to lean towards the generation side of the generation-recognition trade-off. They use source datasets that consist of either studio-quality or high-quality speech, facilitating the training of TTS and VC systems and the successful generation of spoofed speech samples. However, both the bona fide and spoofed speech in these datasets are exceedingly clean, making them far from real-world, noisy speech data.
 
SpoofCeleb is the first dataset to use real-world, noisy, and reverberant data originating from TITW, which originates from VoxCeleb1, as the source for training and synthesizing spoofed speech. We tackle the generation-recognition trade-off by using our carefully curated, fully automated pre-processing pipeline that enables TTS models to be trained on data that more closely mirrors real-world conditions.

\newpara{Datasets for SASV.}
As SASV is an emerging task extending the scope of ASV systems with spoofing robustness, there is a lack of dedicated datasets for SASV. Earlier studies on SASV have relied on SDD datasets~\cite{todisco2018integrated,shim2020integrated}. However, current SDD datasets do not prioritize speaker diversity and balance, both of which are critical for SASV. Most datasets also lack a sufficient number of speakers.

To the best of our knowledge, VSASV~\cite{hoangvsasv}, a parallel data collection effort to SpoofCeleb, is the only attempt at addressing these limitations by creating a dataset specifically for SASV. SpoofCeleb complements VSASV while also having several distinctions. While VSASV includes three spoofing attacks, SpoofCeleb contains 23. Although VSASV uses in-the-wild bona fide data, its spoofed data are derived from high-quality sources due to the challenges in developing TTS systems with in-the-wild data. In contrast, SpoofCeleb adopts TITW which originates from VoxCeleb1, a widely-used ASV dataset recorded in the wild, as its bona fide source. Additionally, VSASV includes approximately 300 k samples, whereas SpoofCeleb offers over 2.5 M samples. Table~\ref{tab:corpora_comparison} compares SpoofCeleb with other SDD and SASV datasets.

\begin{figure}
    \centering
    \includegraphics[width=\columnwidth]{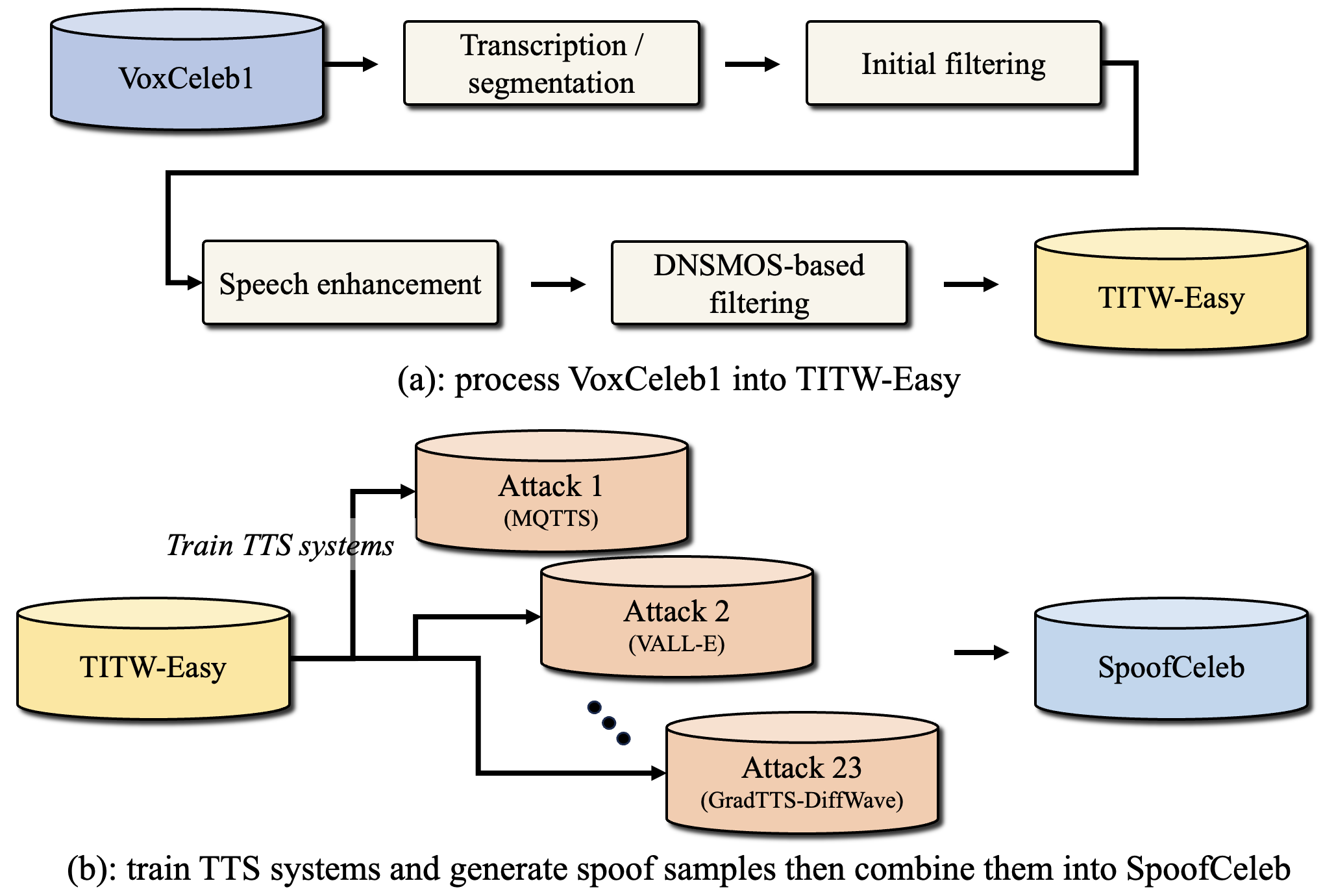}
    \vspace{-0.5cm}
    \caption{
    Overall process pipeline of SpoofCeleb dataset collection.
    (a): our proposed fully automated pipeline transcribes, segments, filters, enhances, and again filters with DNSMOS to derive TITW-Easy~\cite{jung2024text} from VoxCeleb1~\cite{nagrani2017voxceleb}, which is adequate for TTS training.
    (b): 23 different TTS systems are trained using TITW-Easy and spoof speech samples are generated. All generated spoofing samples are combined with TITW-Easy to constitute SpoofCeleb.
    }
    \label{fig:pipeline}
    \vspace{-0.4cm}
\end{figure}

\section{SOURCE DATASET: TITW}
\label{sec:titw}

Our goal is to create a dataset for SDD and SASV using VoxCeleb1 as the source so that both bona fide and spoofed samples would reflect real-world scenarios. However, VoxCeleb1 is not suitable for direct use in TTS training.\footnote{Our preliminary attempts to train TTS systems using the raw VoxCeleb1 data without further processing were unsuccessful.} The challenges with VoxCeleb1 are multifaceted. For example, the speech samples often (i) contain overly emotional expressions, (ii) include extended non-speech segments, or (iii) have excessively long durations.
To address these issues, our developed fully automated pipeline processes VoxCeleb1 into the TITW dataset, which can be used for TTS training.

Figure~\ref{fig:pipeline}-(a) illustrates the automated processing pipeline that was used to generate the TITW dataset. The pipeline begins by transcribing and obtaining word-level alignment using the WhisperX toolkit~\cite{WhisperX-Bain2023}. This toolkit transcribes the speech using the pre-trained Whisper Large v2 Automatic Speech Recognition (ASR) model~\cite{Whisper-Radford2023}, while word-level segmentation is derived from another phoneme-based ASR model. For a small subset of randomly selected samples, we also transcribe the text using the OWSMv3.1 model~\cite{OWSM-Peng2024} and cross-check the accuracy of the transcriptions. We then segment the utterances from VoxCeleb1 whenever a silence longer than 500 ms is detected, resulting in multiple segments from a single utterance.
Next, we apply a series of heuristic-driven rules -- developed through several iterations of TTS training -- to filter the data. We discarded any samples that (i) were non-English, (ii) were shorter than 1 second or longer than 8 seconds, (iii) contained one or more words with a duration exceeding 500ms, or (iv) had empty transcriptions.

After completing the initial processing steps (referred to as TITW-Hard in~\cite{jung2024text}), we conducted multiple iterations of TTS training trials. Despite these efforts, training remained extremely challenging for most TTS systems, with only a few recent models showing success. The generated speech was still insufficient to deceive pre-trained ASV systems, as measured using the SPooF Equal Error Rate (SPF-EER) metric~\cite{jung2022sasv}.\footnote{The SPF-EER is calculated by assessing an ASV system's ability to correctly accept target trials while rejecting spoofed non-target trials. Bona fide non-target trials are excluded from this protocol, as the focus is solely on evaluating the ASV system's spoofing robustness.}
To address this, we applied speech enhancement using a pre-trained model named DEMUCS and excluded samples with DNSMOS ``BAK" (background noisy quality) scores below 3.0. The final number of speech segments (TITW-Easy in~\cite{jung2024text}) is approximately 248 k, which serves as the bona fide portion of the SpoofCeleb dataset. For full details on the preparation of TITW from VoxCeleb1, refer to~\cite{jung2024text}.
Nonetheless, we note that this choice of enhancing the bona fide speech may confuse the training of detection models because inevitable artifacts can be added with the enhancement process. Yet, we employ TITW-Easy as the bona fide, not TITW-Hard, because of the aforementioned practicality.

\section{SPOOFCELEB}
Figure~\ref{fig:pipeline}-(b) illustrates the composition of SpoofCeleb. The TITW dataset serves as the foundation for training multiple TTS systems. These systems are then used to synthesize spoofed speech samples, which are combined with the bona fide speech samples from TITW to form the complete SpoofCeleb. To achieve this, we use 4 acoustic models, 6 waveform models (i.e., vocoders), and 5 End-to-End (E2E) models.
Unless mentioned otherwise, all models were trained from scratch using the TITW-Easy data.
SpoofCeleb does not include voice conversion systems, as TTS systems pose more immediate and prevalent security threats with publicly available APIs. Incorporating voice conversion systems would also require more complex configurations, such as defining source and target speaker pairs. Hence we leave this part for future work.

\subsection{Acoustic models}
Training acoustic models using in-the-wild data was one of the most challenging aspects of SpoofCeleb creation. We applied several criteria to evaluate the success of the training, including (but not limited to) speech intelligibility, measured by the Word Error Rate (WER), noisiness, assessed using DNSMOS, and speaker identity, evaluated using SPF-EER. Among these metrics, SPF-EER was prioritized as the primary measure, since the most critical factor in a spoofing attack is whether it can deceive an ASV system. The final models that were successfully trained include TransformerTTS, GradTTS, Matcha-TTS, and BVAE-TTS.

\newpara{TransformerTTS} ~\cite{li2019transformetTTS} is an autoregressive TTS model that generates mel-spectrograms from textual input using a transformer-based architecture. 
The model uses a sequence of transformer encoder and decoder blocks with multi-head self-attention. 
We trained TransformerTTS using the ESPnet toolkit~\cite{watanabe2018espnet}.\footnote{\url{https://github.com/ESPnet/ESPnet}.}

\newpara{GradTTS.} ~\cite{gradtts2021Popov} is a TTS model with a score-based decoder that generates mel-spectrograms by gradually transforming noise predicted by the text encoder. 
During inference, we set the denoise step to 50 to ensure high-quality speech generation. We used the official implementation and followed the default settings.\footnote{\url{https://github.com/huawei-noah/Speech-Backbones}.}

\newpara{Matcha-TTS.} \cite{mehta2024matcha} is an efficient non-autoregressive TTS model based on an optimal-transport conditional flow matching decoder~\cite{lipmanflow}. 
Unlike score-based models, it constructs a more direct sampling trajectory, enabling high-quality generation with fewer sampling steps. 
We used the official implementation.\footnote{\url{https://github.com/shivammehta25/Matcha-TTS}.}

\newpara{BVAE-TTS.} ~\cite{lee2020bidirectional} uses a Bidirectional-inference Variational AutoEncoder (BVAE) to model the hierarchical relationships between text and speech. 
By leveraging the attention maps generated using BVAE-TTS, the model jointly trains a duration predictor, enabling robust and efficient non-autoregressive speech generation. 
We used the official implementation.\footnote{\url{https://github.com/LEEYOONHYUNG/BVAE-TTS}.}

\subsection{Waveform models}
The training of waveform models was comparatively straightforward.
We employed a mix of both classic and recent waveform models, including  DiffWave, HiFiGAN, Parallel WaveGAN, Neural source-filter model with HiFi-GAN discriminators (NSF-HiFiGAN),
BigVGAN, and WaveGlows.

\newpara{DiffWave.}~\cite{kong2020diffwave} is a diffusion probabilistic model designed for both conditional and unconditional waveform generation. We used the official implementation.\footnote{\url{https://github.com/lmnt-com/diffwave}.}

\newpara{HiFiGAN.}~\cite{NEURIPS2020_c5d73680} is a widely known GAN-based waveform model that uses multiple transposed convolution blocks to progressively upsample and transform input mel-spectrograms into speech waveforms. The generator is optimized using multiple discriminator losses, a feature matching loss, and L1 loss between the generated and ground truth mel-spectrograms. We used the HiFiGAN V1 architecture from the official implementation.\footnote{\url{https://github.com/jik876/hifi-gan}.}

\newpara{Parallel WaveGAN.}~\cite{yamamoto2020parallelwavegan} is a lightweight vocoder model. It uses a non-autoregressive WaveNet~\cite{van2016wavenet} architecture combined with multi-resolution Short-Time Fourier Transform (STFT) loss and waveform adversarial loss. We used the official implementation.\footnote{\url{https://github.com/kan-bayashi/ParallelWaveGAN}.}

\newpara{NSF HiFiGAN.}~\cite{wangNeuralSourceFilterWaveform2020} is similar to Parallel WaveGAN but explicitly incorporates a sine-based source signal as input to the generator. 
It also includes a noise branch that transforms random noise into an aperiodic signal. 
This aperiodic signal is combined with the generator's periodic output for harmonic-plus-noise speech waveform generation. We used the official implementation.\footnote{\url{https://github.com/nii-yamagishilab/project-NN-Pytorch-scripts}.}

\newpara{BigVGAN.}~\cite{lee2022bigvgan} is a universal GAN-based vocoder that generalizes effectively across diverse scenarios, including unseen speakers, languages, and recording environments. 
By using periodic activation functions and anti-aliased representations, BigVGAN introduces a beneficial inductive bias for speech synthesis. 
We used the official implementation\footnote{\url{https://github.com/NVIDIA/BigVGAN}.} 

\newpara{WaveGlow.}~\cite{prenger2018waveglow} generates waveforms through a series of neural network-based invertible affine transformations conditioned on input mel-spectrograms. During training, the model parameters are optimized to whiten the ground-truth waveform as much as possible. We used the same toolkit as with NSF HiFiGAN.

\subsection{E2E and speech-language models with neural codecs}
While two-stage TTS pipelines have proven effective for modeling speech from text, they often suffer from poor quality due to the mismatch between acoustic and waveform models. Waveform models are trained on predefined features but must process the outputs generated by acoustic models during inference, leading to potential inconsistencies. To address this issue, several E2E models have been proposed, and we have successfully trained multiple E2E models using the TITW dataset.

Speech-Language Models (SpeechLMs) represent an emerging category of TTS models. Similar to language models in natural language processing, they are trained to predict tokens, in this case, tokens of neural codecs, which are then decoded via a neural codec system's decoder. Unlike acoustic models, which can be paired with any compatible waveform model, SpeechLMs rely on a predetermined decoder based on the neural codec used during training, limiting their ability to function with multiple decoders.

\newpara{VALL-E, Multi-Scale Transformer, and Delay}. 
VALL-E~\cite{valle} predicts the first token of each frame using an autoregressive module, followed by a non-autoregressive prediction for the remaining tokens. Multi-Scale Transformer~\cite{mst} uses a global Transformer for inter-frame modeling and a local Transformer for intra-frame modeling, maintaining full autoregression without approximation. In Delay~\cite{delay}, the multi-stream token sequences are processed using a ``delay" interleave pattern, which enables approximate autoregressive prediction for both inter- and intra-frame modeling, achieving high efficiency.
We used implementations of the three models in the ESPnet toolkit.$^4$

\newpara{MQTTS.}~\cite{chen2023vector} is designed to synthesize speech using real-world data from YouTube and podcasts. To address misalignments common in mel-spectrogram-based autoregressive models, it uses a multi-codebook vector quantization approach to improve both speech intelligibility and diversity. MQTTS aligns closely with the goals of this work, as we aim to develop a dataset that spans real-world data for both bona fide and spoofed speech. We used the official implementation.\footnote{\url{https://github.com/b04901014/MQTTS}.}

\newpara{VITS.}~\cite{kim2021vits} is an E2E TTS model that combines a conditional VAE with stochastic duration prediction to generate waveforms from textual input. The model uses normalizing flow to learn latent representations from speech, while the stochastic duration predictor captures diverse speech prosody from text. For waveform generation, adversarial loss is used to produce high-quality waveforms from the latent representations. We trained VITS using the ESPnet toolkit.$^4$

\subsection{Attack generation, partitioning, and protocols}
Diverse combinations of acoustic and waveform models, alongside E2E and SpeechLM models, result in a total of 23 spoofing attacks. This approach is inspired by previous research, which demonstrated that both acoustic and waveform models impact the perceptual quality of synthesized speech~\cite{watts19_ssw}. Table~\ref{tab:attack_spec} provides a detailed overview of the 23 spoofing attacks included in SpoofCeleb.

Data partitioning for SpoofCeleb requires a more sophisticated approach compared to existing ASV or SDD datasets. An SDD dataset only requires the binary bona fide or spoof label, while an ASV dataset focuses on speaker identities. SpoofCeleb, as a dataset for both SDD and SASV, must account for both bona fide/spoof labels and speaker identities simultaneously.

\newpara{Speakers.} For the speaker partitioning, we divide the 1,251 speakers in the bona fide data into three sets: $1,171$ for training, $40$ for validation, and $40$ for evaluation. This ensures that there are no overlapping speakers between any of the sets.

\newpara{Spoofing attacks.} For spoofing attacks, we divide the bona fide data (A00) and the 23 spoofing attacks (A01–A23) as follows. In the training set, 10 attacks (A01 to A10) are combined with the bona fide data. Among these attacks, six are derived from a combination of acoustic and waveform models, while the remaining four originate from E2E and SpeechLM TTS systems.

In the validation set, there are 6 attacks: A06, A07, and A11 to A14, combined with the bona fide data (A00). Attacks A06 and A07 represent known attacks from unknown speakers. Attacks A11 and A12 involve the same architecture as other attacks but differ in model training details. Specifically, A11 is fully trained from scratch using the TITW dataset, while in A02, the decoder was pre-trained. Similarly, A12 is fully trained from scratch on TITW, whereas A04 was pre-trained on LibriSpeechGigaSpeech~\cite{gigaspeech} and the English subset of Multilingual LibriSpeech~\cite{pratap2020mls}, then fine-tuned on TITW.
Attacks A13 and A14 serve as partially known attacks. In A13, the acoustic model (GradTTS) is known, but the waveform model (NSF HiFiGAN) is unknown. Similarly, in A14, the acoustic model (Matcha-TTS) is known, but the waveform model (HiFiGAN) is unknown.

\begin{figure}
    \centering
    \includegraphics[width=\columnwidth]{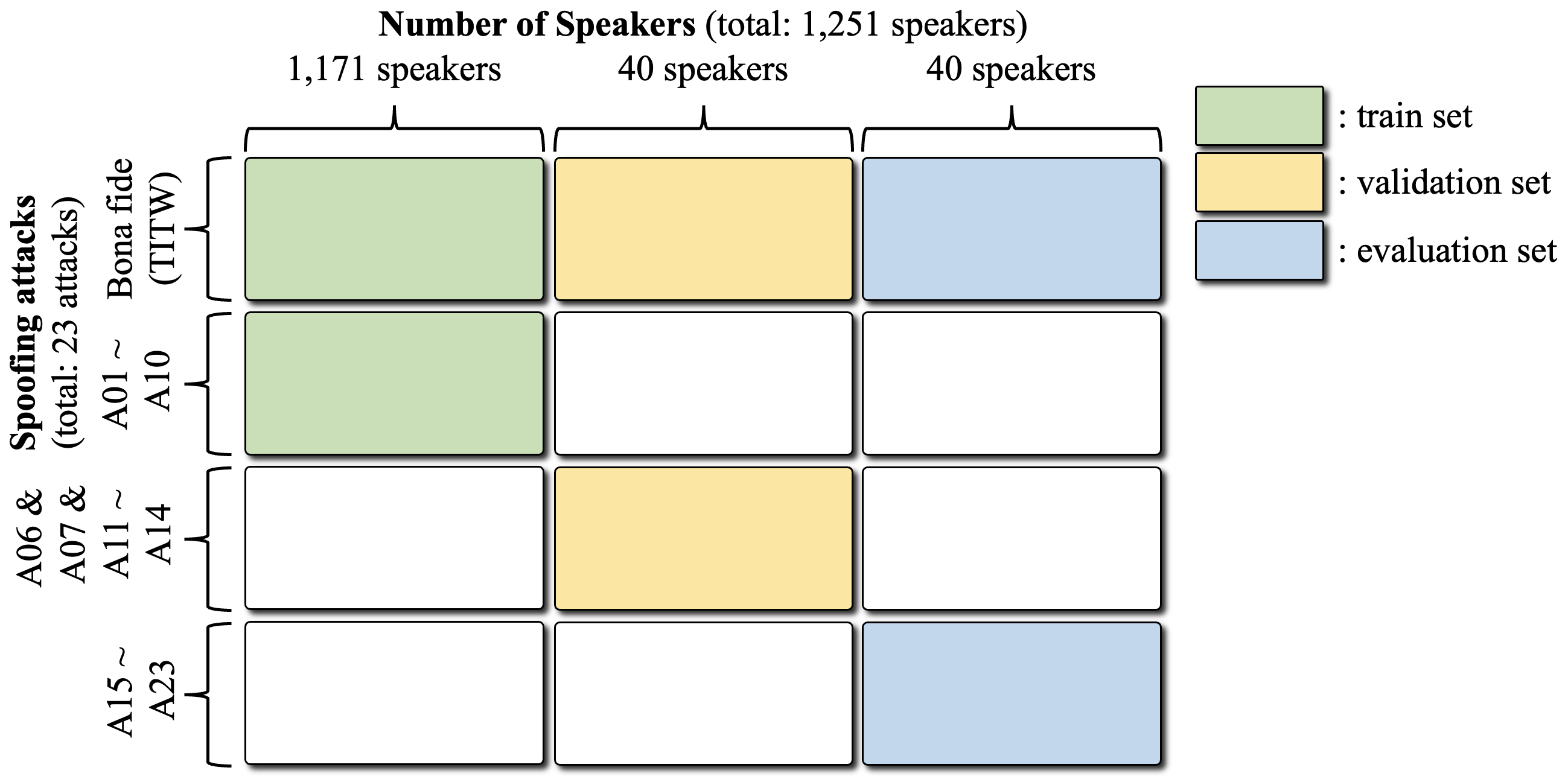}
    \caption{
    Illustration of how SpoofCeleb is partitioned.
    }
    \label{fig:partition}
\end{figure}
\begin{table}[t!]
    \caption{
    Number of speech files and protocols.
    Number of trials equals that of speech files for SDD protocols.
    }
    \label{tab:db_spec}
    \centering
    \begin{tabular}{lcc}
        \toprule
        & \# speech files & \# trials in SASV protocols\\
        \toprule
        Train & $2,540,421$ & N/A\\
        Validation & $55,741$ & $39,353$\\
        Evaluation & $91,130$ & $133,448$\\
        \bottomrule 

    \end{tabular}
    \vspace{-0.3cm}
\end{table}

In the evaluation set, there are 9 attacks, A15 to A23. Attacks A15 and A16 involve known architectures but differ in configurations. For A15, the decoder is initialized with a pre-trained model, and the speaker embeddings are taken from target utterances, simulating a scenario in which an attacker has access to the target speaker's utterance pool. A16 was pre-trained using the same data composition as A04. Attacks A17 and A18 represent partially known attacks where the acoustic models are known, but the waveform models are not. Finally, A19 to A23 are fully unknown attacks, meaning no part of their models was encountered during training.

Figure~\ref{fig:partition} illustrates the three partitions of SpoofCeleb and 
Table~\ref{tab:db_spec} provides the statistics of each partition.
In total, SpoofCeleb contains over 2.5 M speech samples.

\newpara{Protocols.}
SpoofCeleb includes protocols for validating and evaluating developed SDD and SASV models. The SDD protocols for validation and evaluation specify the speech samples to be assessed, while the SASV protocols list pairs of trials with an enrollment utterance and a test utterance. Table~\ref{tab:db_spec} provides details on the number of utterances for the SDD protocols and the number of trials for the SASV protocols.

\begin{table}[t!]
  \caption{
  Spoofing attacks of SpoofCeleb. There are 23 different attacks stemming from 23 different TTS systems. $^\dagger$: pre-trained, $^\ddagger$: decoder is pre-trained, $^\diamond$: speaker embeddings from target utterances.
  }
 \label{tab:attack_spec}
 \centering
 \resizebox{\columnwidth}{!}{
  \begin{tabular}{cccc}
  \toprule
  AttackID & Partition & Acoustic model &  Waveform model\\
  \toprule
  A01 & trn & VITS & N/A\\
  A02 & trn & MQTTS$^\ddagger$ & N/A\\
  A03 & trn & VALL-E & N/A\\
  A04 & trn & Delay$^\dagger$ & N/A\\
  A05 & trn & GradTTS & DiffWave\\
  A06 & trn\&dev & GradTTS & BigVGAN\\
  A07 & trn\&dev & GradTTS & WaveGlow\\
  A08 & trn & MatchaTTS & DiffWave\\
  A09 & trn & MatchaTTS & BigVGAN\\
  A10 & trn & MatchaTTS & WaveGlow\\
  \hline
  A11 & dev & MQTTS & N/A\\
  A12 & dev & Delay & N/A\\
  A13 & dev & GradTTS & NSF HiFiGAN\\
  A14 & dev & MatchaTTS & HiFiGAN\\
  \hline
  A15 & eval & MQTTS$^{\ddagger,\diamond}$ & N/A\\
  A16 & eval & VALL-E$^\dagger$ & N/A\\
  A17 & eval & GradTTS & HiFiGAN\\
  A18 & eval & MatchaTTS & NSF HiFiGAN\\
  A19 & eval & Multi-scale Transformer & N/A\\
  A20 & eval & Multi-scale Transformer$^\dagger$ & N/A\\
  A21 & eval & TransformerTTS & ParallelWaveGAN\\
  A22 & eval & BVAE-TTS & HiFiGAN\\
  A23 & eval & BVAE-TTS & NSF HiFiGAN\\
  \toprule
  \end{tabular}
  }
  \vspace{-0.4cm}
\end{table}

\begin{table*}[t!]
  \caption{
  Quality and strength of 23 spoofing attacks included in SpoofCeleb.
SPF-EER (\%) measures how hard it is to reject an attack by a pre-trained ASV system.
MCD, UTMOS, and DNSMOS demonstrate how noisy the attacks are and WER (\%) measures the intelligibility.
  }
 \label{tab:attack_quality}
 \centering
  \begin{tabular}{ccccccc}
  \toprule
  Attack ID & Partition & SPF-EER (\%)$\downarrow$ & MCD$\downarrow$ & UTMOS$\uparrow$ & DNSMOS$\uparrow$ & WER (\%)$\downarrow$\\
  \toprule
  A00 (bona fide) & trn\&val\&eval & N/A & N/A & 3.32 & 2.78 & 9.10\\
  \hline
  A01 & trn & 29.22 & 8.61 & 2.77 & 2.74 & 53.00\\
A02 & trn & 49.47 & 7.09 & 3.08 & 2.83 & 23.60 \\
A03 & trn & 12.51 & 10.85 & 3.28 & 2.93 & 28.50\\
A04 & trn & 20.86 & 10.42 & 3.59 & 2.83 & 4.80\\
A05 & trn & 23.63 & 6.76 & 2.18 & 2.39 & 11.90\\
A06 & trn\&val & 29.42 & 9.23 & 2.08 & 2.16 & 11.30\\
A07 & trn\&val & 24.61 & 5.61 & 1.30 & 1.51 & 11.90\\
A08 & trn & 32.00 & 5.36 & 2.47 & 2.59 & 15.80\\
A09 & trn & 31.07 & 9.10 & 2.38 & 2.48 & 15.90\\
A10 & trn & 26.20 & 5.66 & 1.32 & 1.79 & 15.70\\
\hline
A11 & val & 47.78 & 6.99 & 3.08 & 2.83 & 23.30\\
A12 & val & 14.52 & 10.91 & 3.26 & 2.84 & 32.50\\
A13 & val & 27.13 & 5.52 & 1.97 & 2.13 & 12.90\\
A14 & val & 29.36 & 5.11 & 2.52 & 2.48 & 14.50\\
\hline
A15 & eval & 65.21 & 6.79 & 3.14 & 2.83 & 21.20\\
A16 & eval & 21.63 & 10.43 & 3.87 & 2.93 & 3.40\\
A17 & eval & 30.20 & 5.44 & 2.62 & 2.43 & 11.20\\
A18 & eval & 25.84 & 5.19 & 2.04 & 2.24 & 16.00\\
A19 & eval & 17.20 & 10.69 & 3.29 & 2.88 & 11.80\\
A20 & eval & 22.36 & 10.73 & 3.53 & 2.92 & 5.50\\
A21 & eval & 22.32 & 11.68 & 2.06 & 2.50 & 24.90\\
A22 & eval & 5.65 & 5.74 & 1.37 & 1.62 & 21.50\\
A23 & eval & 6.75 & 5.65 & 1.30 & 1.50 & 25.90\\
  \toprule
  \end{tabular}
  \vspace{-0.5cm}
\end{table*}

\section{Baselines}
\subsection{SDD}
Two E2E SDD models, RawNet2~\cite{tak2021end} and AASIST~\cite{jung2022aasist}, are used as the baselines. The RawNet2 model for SDD is an adapted version of RawNet2 originally designed for ASV. It features an input layer that processes raw waveforms directly and uses convolution-based residual blocks. Frame-level representations are aggregated, projected, then passed through a binary classification head.

AASIST is one of the most widely used SDD models in recent literature. Like RawNet2, it includes an input layer that processes raw waveforms and uses convolution-based residual blocks. However, unlike RawNet2, AASIST incorporates graph attention network-based modules designed to capture spectral and temporal spoofing artifacts separately. It then uses heterogeneous stacking of graph attention layers to jointly model spectral and temporal information concurrently.

\subsection{SASV}
We employ three models as SASV baselines, all of which use the SKA-TDNN architecture~\cite{mun2023frequency}. These models are used to assess the impact of different training data and scenarios. SKA-TDNN is a convolution-based model with residual connections, incorporating dedicated modules and architectural design choices for multi-scale processing. It is an advanced version of the ECAPA-TDNN architecture~\cite{desplanques2020ecapa}.

Among the three SASV baselines, the first model (``Conventional ASV") is trained as a conventional ASV system using the VoxCeleb1\&2 datasets, without considering spoof robustness. We use a pre-trained model from ESPnet-SPK~\cite{jung2024espnet}. The second model (``SASV trained on out-of-domain data") is trained as an SASV model but uses out-of-domain data from the ASVspoof2019 logical access dataset~\cite{wang2020asvs2019}. We use a pre-trained model from~\cite{mun2023towards}. The third model (``SASV trained on SpoofCeleb") is trained as an SASV model using the training set from SpoofCeleb.

\section{Metrics}
A diverse set of metrics is employed to evaluate the SpoofCeleb dataset, as well as the SDD and SASV models. To assess the quality of the speech samples and the strength of the attacks, we use SPF-EER, Mean Cepstral Distortion (MCD), UTMOS~\cite{saeki22c_interspeech}, DNSMOS~\cite{dnsmos}, and Word Error Rate (WER), with the WER evaluated using the OpenAI Whisper-Large model~\cite{radford2023robust}.
SPF-EER measures speaker characteristics, UTMOS and DNSMOS are objective approximations of perceived quality and noisiness of synthesized speech, and WER measures intelligibility.
For evaluating the performances of the SDD baselines, we use Equal Error Rate (EER) and the min Detection Cost Function (minDCF)~\cite{nistNIST2016}. To assess the SASV baselines, we adopt the recently proposed architecture-agnostic Detection Cost Function (min a-DCF)~\cite{shim2024dcf}, along with Speaker Verification EER (SV-EER) and SPooF EER (SPF-EER). Table~\ref{tab:sasv_metric} outlines the trial types involved in the SASV metrics; a-DCF includes all three trial types, while SV-EER and SPF-EER cover only a subset. 

\begin{table}[t!]
    \caption{
    Three metrics used for gauging performances of SASV baselines.
    a-DCF measures the overall performance.
    SV-EER measures the ability to reject non-target speakers and SPF-EER measures spoof-robustness.
    ``+": a system should accept, ``-'': a system should reject.
    }
    \label{tab:sasv_metric}
    \centering
    \begin{tabular}{lccc}
        \toprule
        Trial type $\backslash$ metric & a-DCF~\cite{shim2024dcf} & SV-EER & SPF-EER~\cite{jung2022sasv}\\
        \toprule
        Target & + & + & +\\
        Bona fide non-target & - & - & \\
        Spoof non-target & - & & -\\
        \bottomrule 

    \end{tabular}
    \vspace{-0.3cm}
\end{table}

\begin{table}[t!]
    \caption{
    SDD baseline performances.
    }
    \label{tab:sdd_baseline}
    \centering
    \begin{tabular}{lcccccc}
        \toprule
        \multirow{2}{*}{System} & \multirow{2}{*}{Train set} & \multicolumn{2}{c}{Validation} & \multicolumn{2}{c}{Evaluation}\\
        & & EER & minDCF & EER & minDCF\\
        \toprule
        RawNet2 & ASVspoof2019 & 	56.33& 0.9996& 	58.79 & 0.9990\\
        AASIST & ASVspoof2019 & 26.64& 0.6048& 23.51 & 0.4710\\
        RawNet2 & SpoofCeleb & 8.63 & 0.1910 & 1.12 & 0.0290\\
        AASIST & SpoofCeleb & 0.61 & 0.0160 & 	2.37& 0.0328\\
        \bottomrule
    \end{tabular}
    \vspace{-0.5cm}
\end{table}
\begin{table*}[t!]
    \caption{
    Attack-wise performance of RawNet2 SDD baseline on validation and evaluation sets.
    Performances reported using EER (\%).
    }
    \label{tab:sdd_attackwise}
    \centering
    \begin{tabular}{l|cccccc|ccccccccc}
        \toprule
        System & A06 & A07 & A11 & A12 & A13 & A14 & A15 & A16 & A17 & A18 & A19 & A20 & A21 & A22 & A23\\
        \toprule
        RawNet2~\cite{tak2021end} & 0.69 & 0.40 & 19.55 & 0.97 & 6.54 & 1.35 & 0.16 & 0.27 & 0.28 & 2.72 & 0.53 & 0.63 & 1.40 & 0.01 & 0.36\\
        \bottomrule

    \end{tabular}
\end{table*}
\begin{table*}[t!]
    \caption{
    SASV baseline performances.
    SKA-TDNN~\cite{mun2023frequency} model architecture is employed.
    Three models are trained in different scenarios. 
    ``Conventional ASV'' is trained with VoxCelebs1\&2~\cite{nagrani2017voxceleb,chung2018voxceleb2} and ``SASV trained on out-of-domain data'' is trained on ASVspoof2019 logical access~\cite{wang2020asvs2019}.
    }
    \label{tab:sasv_baseline}
    \centering
    \begin{tabular}{lcccccc}
        \toprule
        \multirow{2}{*}{System} & \multicolumn{3}{c}{Validation} & \multicolumn{3}{c}{Evaluation}\\
        & a-DCF & SV-EER & SPF-EER & a-DCF & SV-EER & SPF-EER\\
        \toprule
        Conventional ASV~\cite{jung2024espnet} & 0.4973 & 2.55& 23.62& 0.4923 &3.84 & 23.44\\
        SASV trained on out-of-domain data~\cite{mun2023towards} & 0.2901& 33.67 & 55.01 & 0.9998 & 38.94 & 52.24\\
        SASV trained on SpoofCeleb & 
 	
0.3101 & 41.96	& 64.50& 0.2902& 12.78 & 5.00\\
        \bottomrule

    \end{tabular}
    \vspace{-0.4cm}
\end{table*}

\section{Results}
\subsection{Spoofing attacks}
Table~\ref{tab:attack_quality} presents various metrics to assess the speech quality of the 23 synthesized spoofing attacks and how effectively they threaten ASV systems. SPF-EER is the most critical metric, as it measures the extent to which the generated attacks can deceive existing ASV systems. We evaluated SPF-EER using a pre-trained RawNet3 model~\cite{jung2022pushing}, which is publicly available through ESPnet-SPK~\cite{jung2024espnet}.

In the top row, the speech quality evaluations for A00 (bona fide speech) are provided as reference values. The results confirm that the spoofing attacks in SpoofCeleb are highly threatening, with most attacks achieving an SPF-EER over 20\%. The majority of attacks exhibit relatively minor degradation in UTMOS and DNSMOS, indicating the high quality of the synthesized speech samples. Intelligibility, measured using the WER, shows that for most attacks, there is no more than a 10\% deterioration in performance.

\subsection{SDD} Table~\ref{tab:sdd_baseline} presents the results of four baseline SDD systems. We evaluate two SDD models, RawNet2 and AASIST, trained on two different datasets. The models trained on the ASVspoof2019 logical access dataset are used to assess the zero-shot performance on validation and evaluation SDD protocols of SpoofCeleb. The other two models demonstrate the performance of systems trained on in-domain SpoofCeleb training data.

The zero-shot results in the top two rows indicate that existing SDD models not trained on in-the-wild data struggle to distinguish between spoofed samples and bona fide speech. As shown in rows 3 and 4, there is a significant performance improvement when these models are trained using the SpoofCeleb training set, highlighting the importance of training SDD models on in-the-wild data. However, the RawNet2’s result in row 3 is unexpected, as it shows better performance on the evaluation set than on the validation set, while the evaluation set includes totally unknown attacks. To further investigate this, we conduct an analysis of the attack-wise results.

Table~\ref{tab:sdd_attackwise} presents the attack-wise performance of the RawNet2 baseline SDD model trained on the SpoofCeleb training set. Attacks A06 and A07 are classified as known attacks. Attacks A11 to A18 are partially unknown; in these cases, either the acoustic or waveform model is known, or the architecture is familiar but trained with a different configuration. Attacks A19 to A23 represent entirely unknown attacks.

We found that the inferior performance on attack A11 contributed to the validation set results being worse than those on the evaluation set. Interestingly, when comparing A11 and A15, attack A15 is more difficult to distinguish for a conventional ASV system that does not account for spoofing, with SPF-EER values of 47.78\% for A11 and 65.21\% for A15. Both attacks originate from MQTTS; however, A11 was trained entirely from scratch, while A15 utilized a pre-trained decoder. Once an SASV system is trained on the SpoofCeleb training data, A11 becomes more challenging to detect. A deeper investigation into the reasons behind this phenomenon is left for future work.

The comparative analysis in Tables~\ref{tab:attack_quality} and \ref{tab:sdd_attackwise} reveals a discrepancy between the rankings of attacks' SPF-EER on the pre-trained ASV system trained with VoxCeleb and the rankings of attacks' EER on the SDD system trained with SpoofCeleb. 
This divergence may be attributed to the differences in training data, whether the models were trained on SpoofCeleb. The discrepancy could be a result of the fundamental differences in the tasks themselves, as SDD and SASV systems are optimized for distinct objectives.

\subsection{SASV}
Table~\ref{tab:sasv_baseline} presents the performances of three SASV baselines on the SpoofCeleb validation and evaluation protocols. Min a-DCF assesses the overall performance, while SV-EER and SPF-EER evaluate the systems' ability to reject bona fide and spoof non-target trials, respectively.

As expected, a conventional ASV system that does not account for spoof attacks, shown in the first row, fails to reject synthesized speech samples, with an a-DCF exceeding 0.49 on both the validation and evaluation sets. However, it performs well at rejecting bona fide non-target trials.
The results in the second row indicate an improvement in a-DCF for the validation set, but even worse performance on the evaluation set. Both SV-EER and SPF-EER remain very high, indicating that the system trained for SASV with out-of-domain data struggles to reject both types of non-target trials.
The a-DCF of 0.9998 also signifies that the model fails to find an operating point where it can reject both types of non-target trials.
Finally, when trained on the SpoofCeleb training data, the a-DCF on the evaluation set drops to its lowest value (0.2902), and both SV-EER and SPF-EER are more balanced compared with row 1, where the system was only capable of rejecting bona fide non-target trials.

\vspace{-5pt}
\section{Conclusion and remarks}
This paper introduces SpoofCeleb, a dataset for SDD and SASV based on in-the-wild data. To create a dataset that incorporates real-world conditions, we used a fully automated pipeline to process the VoxCeleb1 dataset, making it possible to use it for training TTS systems. 
We further trained 23 TTS systems, partitioning TITW and the TTS systems into SpoofCeleb, which includes training, validation, and evaluation sets. Protocols were defined to train and test both SDD and SASV models, and baseline systems for SDD and SASV were established, trained, and evaluated.

While there are numerous SDD datasets, many are limited in scale or speaker diversity, which has hindered research on single SASV models. We hope SpoofCeleb will serve as the first dataset with enough data to effectively train single SASV systems. Yet, SpoofCeleb has its limitations. In the experiments, some spoofing attacks are shown to be less challenging, as the wild nature of the TITW data complicates the training of robust TTS systems. Future work will focus on advancing TTS training techniques that can better leverage this challenging in-the-wild data.
\vspace{-10pt}

\bibliographystyle{IEEEtran}
\bibliography{refs}

\end{document}